\documentclass[10pt]{iopart}
\usepackage{iopams}
\usepackage{epsfig,cite}
\newcommand{\be}{\begin{equation}}
\newcommand{\ee}{\end{equation}}
\newcommand{\bd}{\begin{displaymath}}
\newcommand{\ed}{\end{displaymath}}
\newcommand{\BE}{\begin{eqnarray}}
\newcommand{\EE}{\end{eqnarray}}

\newcommand{\sgn}{{\rm sgn}}
\newcommand{\erf}{{\rm erf}}
\newcommand{\id}{{\rm 1\!\! I}}

\newcommand{\bq}{\ensuremath{\mathbf{q}}}

\newcommand{\avg}[1]{\left\langle{#1}\right\rangle}
\newcommand{\davg}[1]{\left\langle\!\left\langle{#1}\right\rangle\!\right\rangle}

\begin{document}

\title{On the transition to efficiency in Minority Games}

\author{Tobias Galla\dag~ and Andrea De Martino\ddag}

\address{
\dag The University of Manchester, School of Physics and Astronomy, Schuster Building, Manchester M13 9PL, United Kingdom\\
\ddag CNR/INFM SMC and Dipartimento di Fisica,
  Universit\`a di Roma ``La Sapienza'', p.le A. Moro 2, 00185 Roma, Italy}

\begin{abstract}
The existence of a phase transition with diverging susceptibility in batch Minority Games (MGs) is the mark of informationally efficient regimes and is linked to the specifics of the agents' learning rules. Here we study how the standard scenario is affected in a mixed population game in which agents with the `optimal' learning rule (i.e. the one leading to efficiency) coexist with ones whose adaptive dynamics is sub-optimal. Our generic finding is that any non-vanishing intensive fraction of optimal agents guarantees the existence of an efficient phase. Specifically, we calculate the dependence of the critical point on the fraction $q$ of `optimal' agents focusing our analysis on three cases: MGs with market impact correction, grand-canonical MGs and MGs with heterogeneous comfort levels.

\end{abstract}


\ead{\tt Tobias.Galla@manchester.ac.uk, Andrea.DeMartino@roma1.infn.it}

\section{Introduction}

It is now well known that Minority Games (MGs) can display two types of phase
transitions separating ergodic phases from a non-ergodic regimes
\cite{book1,book2,rev}. One type of transition is characterized by a diverging susceptibility signalling the 
existence of an informationally efficient phase with vanishing
predictability of the bid time-series \cite{prl,hc}. The second type
of transition (referred to as memory-onset transition) occurs
instead at finite integrated response and is marked by a de Almeida-Thouless-instability and a replica-symmetry broken phase at non-zero predictability, similar to phase transitions observed in models of spin glasses \cite{impact1,impact2}.

From a physical viewpoint, the divergence of static and dynamic
susceptibilities signals a sensitivity of the MG-dynamics to
perturbations in the stationary state. A geometric interpretation of
this phenomenon has been devised in\cite{continuum}, and rests on the
observation that the microscopic $N$-dimensional state vector
describing the system of $N$ interacting agents evolves in an $\alpha
N$-dimensional space spanned by the quenched disorder of the
problem. Here $\alpha$ is the key control parameter of the model. The
analysis of the MG shows that the effective dimension of phase space
is reduced to $(1-\phi)N$ as a fraction $\phi\equiv\phi(\alpha)$ of
agents `freezes' during the course of the dynamics: frozen agents are
those who use just one of their strategies in the long run, so that
their degrees of freedom are effectively removed from the dynamics.

The breakdown of ergodicity is observed to occur at some
$\alpha=\alpha_c$ satisfying the condition
$[1-\phi(\alpha_c)]N=\alpha_c N$, i.e. it occurs when the dimensionality of
the reduced phase space becomes equal to that of the space in which the
effective dynamics is defined.

The aim of the present paper is to test variations of the MG, in which
the above geometric picture is systematically modified, for the
existence or otherwise of phase transitions at diverging integrated
response and for the presence of efficient phases associated with this
type of transition. Such alterations of the model occur naturally when
an additional degree of heterogeneity (besides the quenched randomness
of the strategies) is added to the agents' learning rules, and appear
to be a key ingredient of more realistic models of the learning of
agents. They are indeed very much in the spirit of David Sherrington's
approach to Minority Games. David suggested the addition of
heterogeneity and complexity, and to study their effects on the phase
behaviour of the MG in numerous discussions as well as in earlier
joint articles with the authors of the present paper \cite{GallSher05,
multistrat}, and it is a pleasure to submit work along this line to
the special issue in honour of David's 65th birthday.

Specifically, we will here consider MGs with agent-dependent impact
correction, grand-canonical MGs \cite{gcmg} with heterogeneous incentives to trade, and 
El-Farol type games with heterogeneous comfort levels \cite{elfarol1,hetcl}. We
demonstrate that in order to observe a transition to an efficient
phase, it is for a large class of MGs necessary, and in absence of
memory onset transitions also sufficient, that the above geometric
interpretation holds for a {\em finite} fraction of the agents. In
turn, no efficient phase occurs when no such group of agents
exists.

\section{Definitions and general remarks on the efficient-inefficient transition}

Batch MGs \cite{hc} are discrete zero-temperature dynamical systems describing the coupled time-evolution of $N$ agents, labelled by $i=1,\dots,N$ in the following.  Each agent, in the simplest setup, is described by one continuous dynamical variable $q_i(t)$ which evolves 
according to the following rules:
\BE
q_i(t+1)=q_i(t)-\frac{2}{\sqrt{N}}\sum_{\mu=1}^{\alpha N} \xi_i^\mu A^\mu[\bq(t)]+h_i(t),\nonumber\\
A^\mu[\bq(t)]=\frac{1}{\sqrt{N}}\sum_{j=1}^N\omega_j^\mu+\frac{1}{\sqrt{N}}\sum_{j=1}^N
\xi_j^\mu s_j(t),\label{eq:batch0}\\
s_i(t)={\rm sign}[q_i(t)]\nonumber.
\EE
$\alpha$ is here a finite positive control parameter while
$\{\xi_i^\mu,\omega_i^\mu\}$ are quenched random variables, usually
drawn independently from the set $\{-1,0,1\}$ with weights
$1/4,1/2,1/4$, satisfying $\omega_i^\mu\xi_i^\mu=0$. 
$h_i(t)$ is an external perturbation field, used to
measure the response of the system, and will be set to zero
eventually. These equations can be derived from a finance-inspired
setup that has been discussed at length in the literature (see
e.g. \cite{book2}) and we shall not repeat it here in detail. In a
nutshell, agents' choices are encoded in the Ising spins $s_i(t)$,
whose possible values represent the two (heterogeneous) trading
strategies of which every agent disposes. $q_i(t)$ is then a
`valuation' by which agent $i$ assesses the performance of his
strategies, so that if $q_i(t)\to\pm\infty$ asymptotically, then agent
$i$ will stick to one of his strategies ($s_i(t)\to \pm 1$) when
$t\to\infty$. Otherwise, he will keep switching strategies forever.
These two types of agents are referred to as `frozen' and `fickle',
respectively. Note that frozen agents are insensitive to (small)
perturbations of the dynamics in the steady state. $\mu$ denotes the
state of the world and may take on $P=\alpha N$ values. The quantity
$A^\mu$ represents in turn the bid imbalance (`excess demand') in
state $\mu$. Efficient states are characterized by zero bid imbalance
(on average) and zero predictability. In an efficient phase no
statistical forecast of a bid imbalance is possible in any state
$\mu$. The predictability of the system is measured by the quantity
$H=(\alpha N)^{-1}\sum_\mu
\avg{A|\mu}^2$, with $\avg{\cdot|\mu}$ a time-average conditioned on the occurrence of information pattern $\mu$. The condition $H=0$ implies $\avg{A|\mu}=0$ for all
$\mu=1,\dots,\alpha N$, while $H>0$ signals the presence of an
exploitable pocket of predictability in the time series of bid
imbalances.

The dynamic update rules (\ref{eq:batch0}) demonstrate that, in
absence of perturbations $h_i(t)$, the state vector
$\bq(t)=(q_1(t),\dots,q_N(t))$ moves in the space spanned by the
$\alpha N$ $N$-dimensional vectors $\bxi^1\,\dots,\bxi^{\alpha N}$. A
breakdown of the ability to remove dynamical perturbations hence
occurs when the space of possible perturbations assumes a higher
dimensionality than $\alpha N$. Due to the freezing of $\phi(\alpha)N$
agents, $[1-\phi(\alpha)]N$ linearly independent modes of perturbations can be applied. The condition for ergodicity hence reads $1-\phi(\alpha)<\alpha$ so that ergodicity breaking occurs at a value of $\alpha$ such that $\alpha_c=1-\phi(\alpha_c)$. 

The relation of this type of transition with the existence of a fully
efficient phase, in which $H=0$, can be understood as follows: as
discussed above $H=0$ implies $P=\alpha N$ conditions, one on each
$\avg{A|\mu}$. With $(1-\phi)N$ effective degrees of freedom available
the system is able to evolve into an asymptotic stationary state $H=0$
at most when $1-\phi>\alpha$, i.e. in the phase where ergodicity is
broken. Hence, the onset of a divergence of the integrated response
function occurs precisely at the phase boundary separating a fully
efficient ($H=0$) non-ergodic phase from an inefficient ($H>0$) ergodic
one. It is here worth noting that, although $H$ is not a strict Lyapunov function of  the MG dynamics \cite{book1,book2}, the pseudo-Hamiltonian $H$ is effectively minimised in the stationary states of the dynamics. This observation in fact allows for equilibrium approaches to the MG, based on the replica method. We will not pursue these in the present paper, however, but restrict to dynamical analyses using generating functional techniques \cite{book2}.

The geometric picture of $\bq$ moving in the space spanned by the $\bxi^\mu$'s is violated, whenever the bid imbalance is agent-specific, i.e. when
\be
q_i(t+1)=q_i(t)-\frac{2}{\sqrt{N}}\sum_{\mu=1}^P \xi_i^\mu A_i^\mu[\bq(t)],
\ee
where $A_i^\mu$ now carries an explicit index $i$. This is the case in MGs
in which agents correct for their own impact on the global bid
\cite{MCZ}, in dilute MGs \cite{dilut} and in El-Farol games with
heterogeneous comfort levels \cite{elfarol1,hetcl}. The update rule of
grand-canonical Minority Games (GCMGs) can also be captured by
defining a suitable agent-specific bid, as detailed below. Indeed the
susceptibility remains finite and $H>0$ for all $\alpha>0$ in such
games.

One may here think of the following extreme cases: the standard MG, $A_i^\mu$ is fully independent of $i$, i.e one has $A_i^\mu=A^\mu$ for all $i$. The opposite case corresponds to full heterogeneity, i.e. $A_i^\mu\neq A_j^\mu$ with probability one for any $i\neq j$. The purpose of the present paper is to study intermediate cases, i.e. MG-type games in which there is some heterogeneity in the update rules, but where groups of extensive (${\cal O}(N)$) size use the same bid imbalance to update their scores. To this end we will in the following introduce a parameter $0\leq q \leq 1$, and study models in which the above geometric picture hold for a group of $qN$ agents (who all use the same bid-imbalance to update their score valuations), but where the remaining $(1-q)N$ agents display heterogeneity in their learning rules. $q$ therefore allows to interpolate smoothly between the MG without additional heterogeneity ($q=1$), and the case of fully heterogeneous agents ($q=0$).

\section{Minority Games with agent-specific impact correction}
We will first address MGs with so-called impact correction \cite{impact1,impact2}. Here,
agents take into account the effects of their own trading actions. We will here not give details of the interpretation as these have been discussed at length in the literature \cite{book1,book2}, but will start from 
\be
A_i^\mu[\bq]=\frac{1}{\sqrt{N}}\sum_{j=1}^N\left(\omega_j^\mu+s_j\xi_j^\mu\right)-\frac{\eta_i}{\sqrt{N}}\left(\omega_i^\mu+s_i\xi_i^\mu\right).
\ee
We here assume that the $\{\eta_i\}$ are drawn independently from a distribution $R(\eta_i)$ at the beginning of the game, and that $R(\cdot)$ is identical for all players.

Two well-developed techniques, adapted from spin glass physics and the
theory of disordered systems, are available to study MGs of the type
discussed in this paper. They rest on equilibrium and non-equilibrium
approaches and are based on replica theory and generating functionals
respectively. Both approaches are discussed in the recent textbooks
\cite{book1,book2}. We will here not enter the mathematical details of
the derivation of the resulting effective macroscopic theories. The
generating functional analysis of MGs with heterogeneous impact
correction factors leads to an ensemble of effective stochastic
processes
\be
q_\eta(t+1)=q_\eta(t)-\alpha\sum_{t'} (\id+G)(t,t') s_\eta(t')+\alpha\eta s_\eta(t)+\sqrt{\alpha}\zeta_\eta(t),
\ee
one for each value of $\eta$ in the support of $R(\cdot)$. The relevant macroscopic order parameters are the correlation and response functions, and are given by
\be
C(t,t')=\int d\eta R(\eta) C_\eta(t,t'), ~~~~G(t,t')=\int d\eta R(\eta) G_\eta(t,t'), ~~~
\ee
where
\be
C_\eta(t,t')=\davg{\left. s_\eta(t)s_\eta(t')\right|\eta}, ~~~~ G_\eta(t,t')=\frac{1}{\sqrt{\alpha}}\davg{\left.\frac{\delta s_\eta(t)}{\delta \zeta_\eta(t')}\right|\eta}.~~~
\ee
Here $\davg{\left.\cdot\right|\eta}$ refers to an average over realisations of the effective process for a given value of $\eta$ only. The noise $\zeta_\eta$ has covariance
\be
\davg{\left.\zeta_\eta(t)\zeta_\eta(t')\right|\eta}=[(\id+G)^{-1}D(\id+G^T)^{-1}](t,t')
\ee
where $D(t,t')=1+C(t,t')$ for all $t,t'$, and $\id$ is the identity matrix. We here note that the covariance of $\zeta_\eta$ does not depend on $\eta$, i.e. it is identical over the entire ensemble. We still keep the subscript in $\zeta_\eta$. 

The analysis then proceeds by assuming a time-translation invariant state ($C_\eta(t,t')=C_\eta(t-t'), G_\eta(t,t')=G_\eta(t-t')$) at finite integrated responses
\be
\chi_\eta=\int d\tau ~ G_\eta(\tau)<\infty.
\ee
We will also write $c_\eta$ for the persistent part of the correlation function $C_\eta(t-t')$, and proceed by taking the long-time average of the effective processes:

\be\label{eq:tilde}
\widetilde q_\eta=\sqrt{\alpha}\zeta_\eta-\frac{\alpha s_\eta}{1+\chi}+\alpha\eta s_\eta,
\ee
with $\widetilde q_\eta=\lim_{t\to\infty} q_\eta(t)/t$, and $s_\eta$ and $\zeta_\eta$ the asymptotic time-averages of $s_\eta(t)$ and $\zeta_\eta(t)$. Further details of these steps can be found in \cite{hc,book2}. Note than $\chi$ appearing in the denominator on the RHS of (\ref{eq:tilde}) is $\chi$ and not $\chi_\eta$. The static noise $\zeta_\eta$ has mean zero and variance
\be
\davg{\left.\zeta_\eta^2\right|\eta}=\frac{1+c}{(1+\chi)^2}
\ee
(independently of $\eta$), where
\be\label{eq:cchi}
c=\int d\eta ~R(\eta) ~c_\eta, ~~~~~ \chi=\int d\eta ~R(\eta)~\chi_\eta.
\ee

Defining
\be\label{eq:v}
v(\eta)=\frac{\sqrt{\alpha}(1-\eta(1+\chi))}{\sqrt{2(1+c)}}
\ee
one then finds after separating frozen and fickle agents along the lines of \cite{hc,book2}
\BE\label{eq:ceta}
c_\eta&=&1+\frac{1-2v(\eta)^2}{2v(\eta)^2}\erf[v(\eta)]-\frac{1}{v(\eta)\sqrt{\pi}}e^{-v(\eta)^2}
\EE
for the persistent part of the correlation function $C_\eta$, and
\BE\label{eq:chieta}
\chi_\eta&=&\frac{1+\chi}{\alpha(1-\eta(1+\chi))}\erf[v(\eta)]
\EE
for the integrated response of agents with impact correction factor $\eta$.

Equations (\ref{eq:cchi})--(\ref{eq:chieta}) fully determine the persistent order parameters $c$ and $\chi$. We here note that the predictability $H$ can be shown to be given by
\be
H=\frac{1}{2}\frac{1+c}{(1+\chi)^2},
\ee
(see \cite{book1,book2} for details) so that a divergence of the
integrated response $\chi$ signals the onset of a fully efficient
phase, characterised by $H=0$, as discussed in the previous section.

Since the error function is bounded, one may conclude from Eq. (\ref{eq:chieta}) that $\chi_\eta$ can never diverge for any $\eta\neq 0$ at finite value of $\alpha$. More precisely, let us write 
\be
R(\eta)=q\delta(\eta)+(1-q)P(\eta),
\ee
where $\int d\eta P(\eta)=1$, and where $P(\eta)$ has no mass concentration (e.g. a $\delta$-peak) at $\eta=0$, i.e. where we impose $\lim_{\delta\to 0} \int_{-\delta}^\delta d\eta P(\eta)=0$.
Then using
\BE
\frac{\chi}{1+\chi}=\int d\eta R(\eta)\frac{\erf[v(\eta)]}{\alpha(1-\eta(1+\chi))}
\EE
we have
\BE
\frac{\chi}{1+\chi}=q \frac{\erf[v(0)]}{\alpha}+(1-q)\int d\eta P(\eta)\frac{\erf[v(\eta)]}{\alpha(1-\eta(1+\chi))}.
\EE
If $q=0$ then this reduces to
\BE
\frac{\chi}{1+\chi}=\int d\eta P(\eta)\frac{\erf[v(\eta)]}{\alpha(1-\eta(1+\chi))},
\EE
and in the integral we always have $\eta\neq 0$ (since $P(\eta)$ has no mass at zero). As a consequence $\chi$ can never diverge, since the LHS of this equation approaches unity when $|\chi|\to\infty$, whereas the RHS tends to zero (again note the boundedness of the error function). We conclude that for $q=0$ there can be no $\chi\to\infty$ transition, and hence no efficient phase.

\begin{figure}[t!!!]
  \vspace*{10mm} \epsfxsize=60mm \epsffile{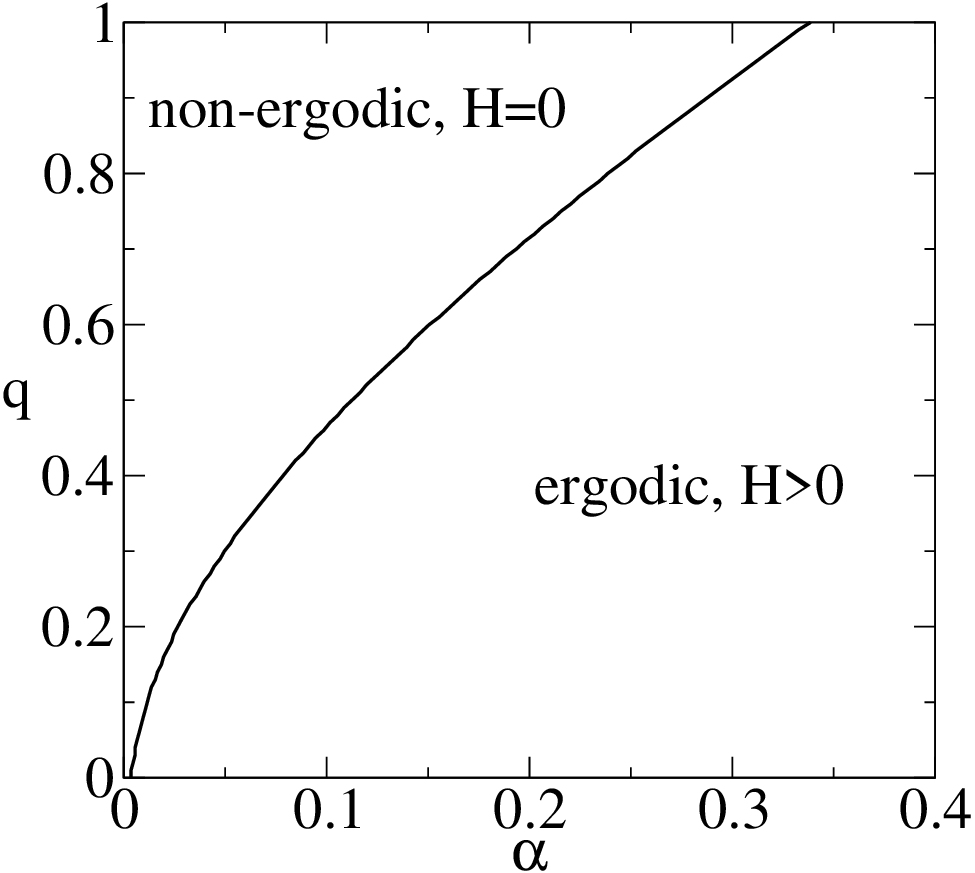} ~~~~
  \epsfxsize=60mm \epsffile{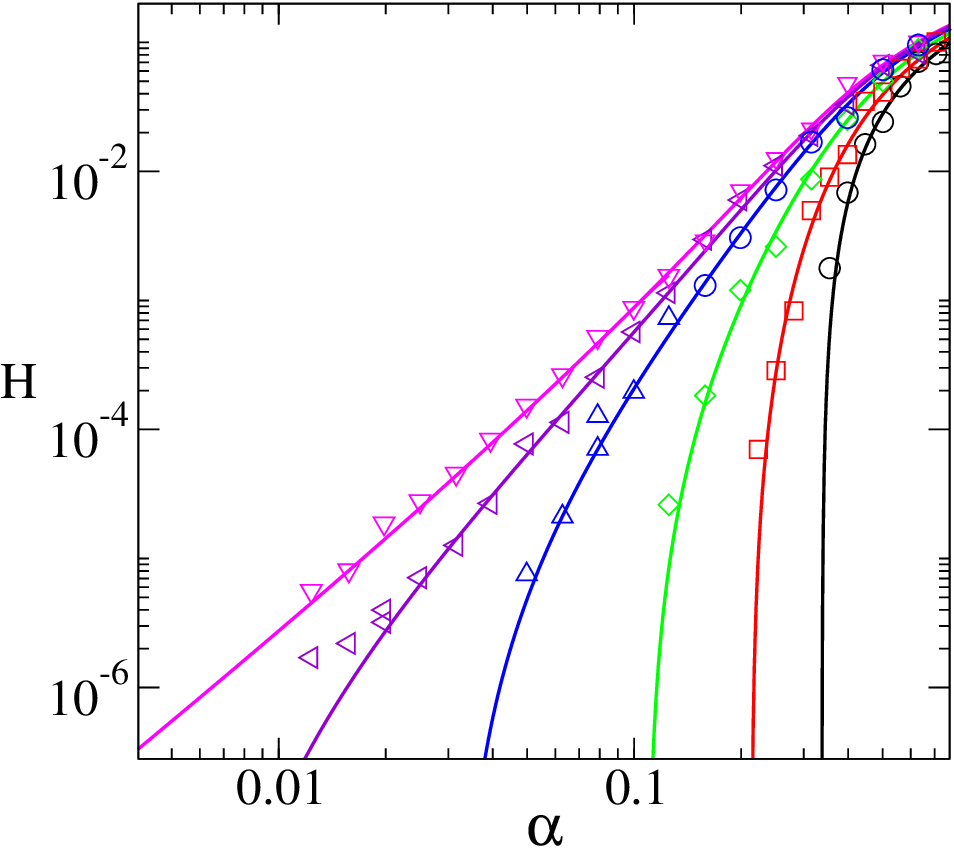} \vspace*{2mm}
  \caption{(Colour on-line) {\bf Left:} Phase diagram for $R(\eta)=q\delta(\eta)+(1-q)\id_{[-1,0]}(\eta)$. {\bf Right:} $H$ versus $\alpha$ for $q=0,0.1,0.25,0.5,0.75,1$ (left to right). Lines are from theory, symbols from simulations ($\alpha N^2=1.6\cdot 10^5$, $1000/\sqrt{\alpha}$ batch iterations, averages over $5$ samples).}
\label{fig:etareta}
\end{figure}

To illustrate these findings we show the phase diagram of the model in
which each $\eta_i$ is drawn from a flat distribution
$P(\eta)=\id_{[-1,0]}$ over the interval $[-1,0]$ with probability
$1-q$, and where $\eta_i=0$ with probability $q$ for each $i$
($\id_{[a,b]}(x)=1$ for $a\leq x\leq b$ and zero otherwise). The left
panel of Fig. \ref{fig:etareta} demonstrates that the integrated
response diverges at a finite value of $\alpha=\alpha_c(q)$ for any
$q>0$, resulting in an efficient phase at $\alpha<\alpha_c(q)$. The
location of the transition $\alpha_c(q)$ tends to zero as $q\to 0$, so
that the transition and efficient phase are absent for $q=0$. In the
right panel of Fig. \ref{fig:etareta} we depict the predictability $H$
in dependence of $\alpha$ for different values of $q$, again
illustrating the fact that a transition between an inefficient and an
efficient phase is present for any $q>0$, but that the phase with
positive predictability persists for all $\alpha>0$ in the case
$q=0$. We here note that simulations at small values of $\alpha$ are
costly in terms of CPU time and that numerical data at $\alpha$ smaller
than approximately $0.02$ may hence be prone to finite-size and
equilibration effects. To conclude this section we here note that we
have explicitly excluded the case $\eta>0$, as it is known that a
memory-onset (MO) transition here precedes the transition marked by a
diverging integrated response \cite{impact1,impact2}. The MO
transition appears to persist when heterogeneous distributions of the
$\{\eta_i\}$ are considered and if at least a finite fraction of
impact correction factors is positive, pre-empting the occurrence of
an efficient phase.

\section{Minority Games with heterogeneous comfort levels}
We next consider MGs with heterogeneous comfort levels, sometimes also referred to as El-Farol games or `clubbing' games  \cite{elfarol1,hetcl}. In such games agents do not apply a strict minority rule when updating their scores (the minority rule dictates that a good move is to make a bid of the opposite sign of the total bid in the market), but apply an individual comfort level $\lambda_i$. Thus the effective bid $A_i$ agent $i$ uses to update his strategy valuations is $A_i(t)=A(t)-\lambda_i$, and we have
\BE
\label{eq:batchcomfort}
\hspace{-2cm}q_i(t+1)=q_i(t)-\frac{2}{\sqrt{N}}\sum_{\mu}\xi_i^{\mu}\left[
\frac{1}{\sqrt{N}}\sum_j\left\{ \xi_j^\mu
s_j(t)+\omega_j^\mu\right\}-\lambda_i\right]+h_i(t).
\EE
Similar to the case of market impact correction we assume that the $\{\lambda_i\}$ are drawn independently from a distribution $R(\lambda_i)$ at the beginning of the game.

The efficient phase here corresponds to a regime in which the population of agents is able to co-ordinate their overall behaviour so that the mean total attendance $\avg{A}$ corresponds (on average) to the mean comfort level.

 The generating functional analysis  again results in an ensemble of effective agent processes, one for each comfort level 
\BE
q_\lambda(t+1)=q_\lambda(t)-\alpha\sum_{t'\leq t}
(\id+G)^{-1}_{tt'}s_\lambda(t')+\sqrt{\alpha}\eta_\lambda(t)+h_\lambda(t).
\EE
$s_\lambda(t)$ is given by $s_{\lambda}(t)=\sgn[q_\lambda(t)]$ and
$\eta_{\lambda}(t)$ is Gaussian noise of zero mean, and with
temporal correlations 
\BE \hspace{-2cm}\davg{\eta_\lambda(t)
  \eta_{\lambda}(t')|\lambda}=[(\id+G)^{-1}D(\id+G^T)^{-1}]_{tt'}+2f_t
f_{t'}-2\lambda(f_t+f_{t'})+2\lambda^2E_{tt'}.  \EE Here
$\id_{tt'}=\delta_{tt'}$ is the identity matrix and $E_{tt'}=1~\forall
t,t'$, $C$ and $G$ are the correlation and response
functions of the system respectively , and we find an additional dynamical order parameter $G'$:
\BE
&C_{tt'}=\lim_{N\to\infty}N^{-1}\sum_i \overline{\davg{s_i(t)s_i(t')}},
~~ G_{tt'}=\lim_{N\to\infty}N^{-1}\sum_i\frac{\partial\overline{\davg{s_i(t)}}}{\partial h_{i}(t')}, \nonumber \\
& G'_{tt'}=\lim_{N\to\infty}N^{-1}\sum_i \lambda_i \frac{\partial
  \overline{\davg{s_i(t)}}}{\partial h_{i}(t')}.
\EE 
The matrix $D$ is, as before, given by $D_{tt'}=1+C_{tt'}$ for all $t,t'$ and we have \BE
f_t=\sum_{t'}[(\id+G)^{-1}G')]_{tt'}. \EE  
These
order parameters are then to be determined as averages over
realisations of the effective processes and over the distribution of
$\lambda$ \BE C_{tt'}&=&\int d\lambda R(\lambda)
\davg{s_\lambda(t)s_\lambda(t')|
  \lambda} \\
G_{tt'}&=&\int d\lambda R(\lambda) \frac{\partial}{\partial
  h_\lambda(t')}\davg{s_\lambda(t)|\lambda} \\
G'_{tt'}&=&\int d\lambda R(\lambda) \lambda \frac{\partial}{\partial
  h_\lambda(t')}\davg{s_\lambda(t)|\lambda} \EE (where
$\davg{\cdots|\lambda}$ is an average over realisations of the
effective process restricted to representative agents with comfort
level $\lambda$).

Assuming time-translation invariance (i.e. $C_{tt'}=C(t-t')$ and
similarly for $G_{tt}$ and $G'_{tt'}$) and finite integrated response
one follows the standard ansatz to proceed from the effective agent
problem to explicit equations characterising the relevant persistent
order parameters of the ergodic stationary states. In our problem
these are given by $c$, the persistent part of the correlation
function, and by $\chi=\sum_\tau G(\tau)$ and $\chi'=\sum_{\tau}
G'(\tau)$. 

The resulting $3\times 3$ system of non-linear equations
for $\{c,\chi,\chi'\}$ then reads
\BE
\hspace{-2cm}c=\int d\lambda ~R(\lambda) ~c_\lambda, ~~~ \chi=\int
d\lambda ~R(\lambda) ~\chi_\lambda, ~~~ \chi'=\int d\lambda
~R(\lambda) \lambda ~\chi_\lambda,
\EE
with $R(\lambda)$ the distribution from which the comfort levels $\{\lambda_i\}$ are drawn and where $\chi_\lambda$
and $c_\lambda$  given by
\BE
\chi_\lambda&=&\frac{(1+\chi)}{\alpha}\erf\left(w(\lambda)\right),\nonumber \\
c_\lambda&=&1+\frac{1-2w(\lambda)^2}{2w(\lambda)^2}\erf\left(w(\lambda)\right)-\frac{1}{w(\lambda)\sqrt{\pi}}e^{-w(\lambda)^2},
\EE
where 
\be
w(\lambda)=\sqrt{\frac{\alpha}{2g(\lambda)(1+\chi^2)}}
\ee
with $g(\lambda)$ is the persistent part of the temporal
correlations of the noise $\eta_\lambda(t)$ in the
effective agent problem:
\be
g(\lambda)=\frac{1+c}{(1+\chi)^2}+2\frac{\chi'^2}{(1+\chi)^2}-4\lambda\frac{
\chi'}{1+\chi}+2\lambda^2\ .
\ee
The mean attendance level comes out as
\be
\avg{A}=\frac{\chi'}{1+\chi}.
\ee

We next perform an analysis similar to the one of the MGs with impact corrections, and look for possible divergences of $\chi$ and $\chi'$. One has
\BE
\frac{\chi}{1+\chi}&=&\alpha^{-1}\int d\lambda R(\lambda)\erf
\left(w(\lambda)\right) 
\EE
and
\BE
\frac{\chi'}{1+\chi}&=&\alpha^{-1}\int d\lambda R(\lambda)\lambda\erf
\left(w(\lambda)\right) 
\EE
For $\chi\to\infty$, we find from the former relation
\BE
1&=&\alpha^{-1}\lim_{\chi\to\infty}\int d\lambda R(\lambda)\erf
\left(w(\lambda)\right),
\EE
 i.e. not all $w(\lambda)$ can go to zero as $\chi\to\infty$. If this happened for all $\lambda$ for which $R(\lambda)$ has mass, then the RHS would go to zero.

Thus, in order for a transition at diverging integrated response to occur, the mass of all $\lambda$ for which  $\lim_{\chi\to\infty}|g(\lambda)(1+\chi)^2|<\infty$ must be positive. Now, we have
\BE
g(\lambda)(1+\chi)^2=1+c+2\chi'^2-4\lambda\chi'(1+\chi)+2\lambda^2(1+\chi)^2.
\EE
Assuming reasonably that $\chi'/\chi\to\gamma$ as $\chi\to\infty$,
with $\gamma$ finite, one finds that $g(\lambda)(1+\chi)^2$ remains
finite, if and only if
\be
\gamma^2+\lambda^2-2\lambda\gamma=0
\ee
i.e. if $\lambda=\gamma$. In other words, if $\chi'/\chi\to\gamma$ as the susceptibilities diverge, then the transition can only exist if $R(\lambda)$ has positive mass concentration (delta-peak) at $\lambda=\gamma$.

Let us take for example a symmetric distribution of $\lambda$ around a value $\Lambda$. Then one has $R(\Lambda-\Delta)=R(\Lambda+\Delta)$. Now $\chi'=\Lambda\chi$ is then a self-consistent solution at the transition, because
\BE
\frac{\chi'}{\chi}=\frac{\int d\lambda R(\lambda)\lambda \erf
\left(\frac{\sqrt{\alpha}}{\sqrt{2g(\lambda)}(1+\chi)}\right)}{\int d\lambda R(\lambda)\erf
\left(\frac{\sqrt{\alpha}}{\sqrt{2g(\lambda)}(1+\chi)}\right)} 
\EE
and $g(\Lambda+\Delta)=g(\Lambda-\Delta)$ for diverging $\chi,\chi'$ and $\chi'/\chi \to \Lambda$.

\begin{figure}[t!!!]
  \vspace*{10mm} \epsfxsize=60mm \epsffile{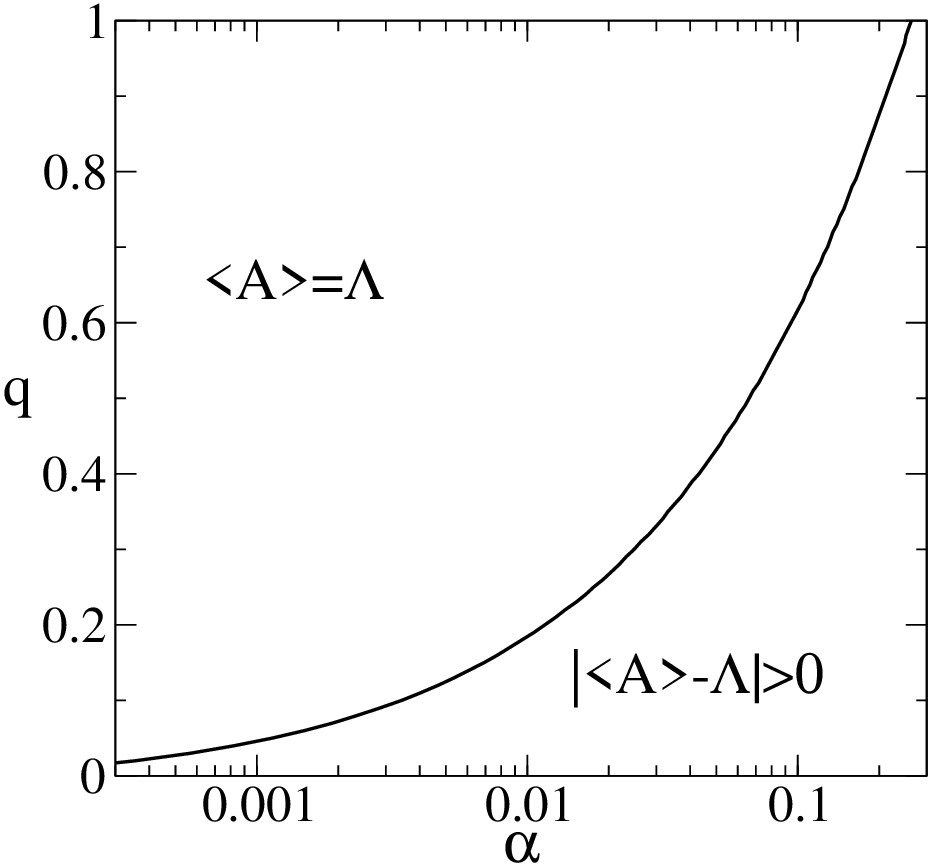} ~~~~
  \epsfxsize=60mm \epsffile{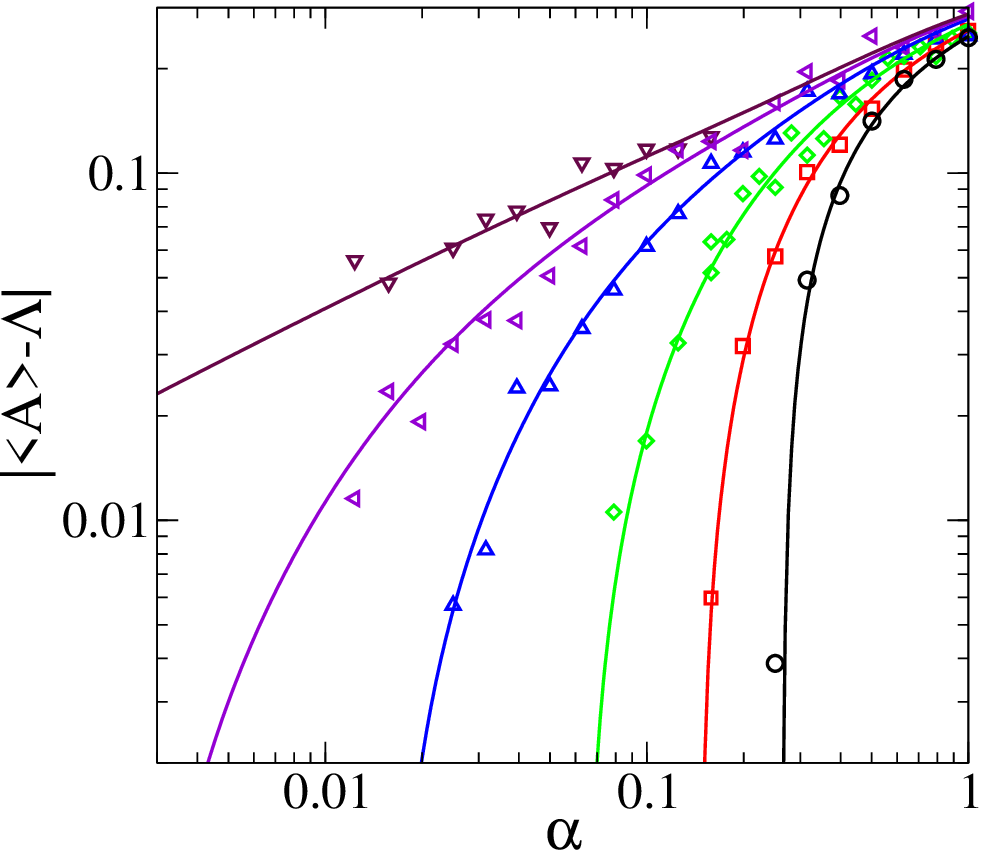} \vspace*{2mm}
  \caption{(Colour on-line) {\bf Left:} Phase diagram for $R(\lambda)=q\delta(\lambda-1/2)+(1-q)\id_{[0,1]}(\lambda)$. {\bf Right:} Deviation of the mean overall bid $\avg{A}$ from the mean comfort level $\Lambda=0.5$ versus $\alpha$ for $q=0,0.1,0.25,0.5,0.75,1$ (left to right). Lines are from theory, symbols from simulations ($\alpha N^2=1.6\cdot 10^5$, $1000/\sqrt{\alpha}$ batch iterations, averages over at least $5$ samples).}
\label{fig:comfort}
\end{figure}

Thus if one has a symmetric distribution of $\lambda$ around a value
$\Lambda$, then the $\chi\to\infty$ transition can exist only if
$R(\lambda)$ has a delta-peak at its mean $\Lambda$ (plus other
symmetric contributions). This is confirmed in Fig. \ref{fig:comfort}
where we depict the resulting phase diagram for
$R(\lambda)=q\delta(\lambda-1/2)+(1-q)\id_{[0,1]}(\lambda)$ at
different values of $q$, and where a phase in which the mean comfort
level $\Lambda=0.5$ is retrieved ($|\avg{A}-\Lambda|=0$) is found for
any $q>0$, but where such a phase is absent at $q=0$.

\section{Grand-canonical Minority Game}

Here we consider the case of grand-canonical MGs \cite{gcmg} with
heterogeneous incentives. In GCMGs agents hold only one active
strategy, but have the option to abstain from taking any action at any
given time step. The batch upate rule for the scores of active
strategies can then be written in the form \cite{finitemem}
\be
q_i(t+1)=q_i(t)-\frac{1}{N}\sum_j\sum_\mu a_i^\mu a_j^\mu n_j(t)-\alpha\varepsilon_i,
\ee
where the $\{a_i^\mu\}$ encode the active strategies of players, and
where the $\{\varepsilon_i\}$ are incentives of agents not to
trade. $n_i(t)$ is given by $n_i(t)=\Theta[q_i(t)]$. One realises that
the geometric interpretation of the score-vector moving in the space
spanned by the disorder holds when $\varepsilon_i\equiv 0$, and in
this case a transition between an efficient and a non-efficient regime
is indeed observed \cite{gcmg,book1,book2}.  The generating functional
analysis for the GCMG is detailed in \cite{book2,finitemem}. The
representative process for agents with incentive $\epsilon$ reads
\be
q_\epsilon(t+1)=q_\epsilon(t)-\alpha\sum_{t'\leq t}(\id+G)(t,t')n_\epsilon(t')-\alpha\epsilon+\sqrt{\alpha}z(t),
\ee
where $n_\epsilon(t)=\Theta[q_\epsilon(t)]$ and
\BE
\avg{z(t)z(t')}=[(\id+G)^{-1}C(\id+G^T)^{-1}](t,t')\\
C(t,t')=\int d\epsilon R(\epsilon)C_\epsilon(t,t')=\int d\epsilon R(\epsilon)\avg{n_\epsilon(t)n_\epsilon(t')}\\
G(t,t')=\int d\epsilon R(\epsilon)G_\epsilon(t,t')=\frac{1}{\sqrt{\alpha}}
\int d\epsilon R(\epsilon)\frac{\partial}{\partial z(t')}\avg{n_\epsilon(t)}
\EE
$\avg{\cdot}$ here denotes an average over the static noise $z$. In the ergodic steady state, the effective noise $z(t)$ becomes a static random variable with variance $\avg{z^2}=c/(1+\chi)^2$, with $c=\int d\epsilon R(\epsilon)c_\epsilon$ and $\chi=\int d\epsilon R(\epsilon)\chi_\epsilon$. Following \cite{finitemem} and introducing the shorthand $\gamma=\sqrt{\alpha}/(1+\chi)$, the persistent autocorrelation and susceptibility at fixed $\epsilon$ are found to be given by
\BE\fl
c_\epsilon=\avg{\theta(z-\gamma-\epsilon\sqrt{\alpha})}+\avg{\frac{(z-\epsilon\sqrt{\alpha})^2}{\gamma^2}\theta(\gamma+\epsilon\sqrt{\alpha}-z)\theta(z-\epsilon\sqrt{\alpha})}\\
\chi_\epsilon=\frac{1}{\gamma\sqrt{\alpha}}\avg{\theta(\gamma+\epsilon\sqrt{\alpha}-z)\theta(z-\epsilon\sqrt{\alpha})}
\EE
The above equations can be solved numerically, and in turn $c$ and $\chi$ can be computed for any distribution of incentives $R(\epsilon)$. As is customary, we divide the $N$ agents in two groups: $N_p$ producers with $\epsilon=-\infty$ (for whom $c_{-\infty}=1$ and $\chi_{-\infty}=0$) and $N_s$ speculators with finite $\epsilon$. We further fix $n_p=N_p/P=1$ and use $n_s=N_s/P$ as the control parameter in place of $\alpha$, which is now given by $\alpha=1/(n_s+n_p)$. For the sake of simplicity, we work out the theory in detail only for the case where $R(\epsilon)$ is symmetric around $\epsilon=0$, namely
\be\label{ere}
R(\epsilon)=q\delta(\epsilon)+\frac{1-q}{2}[\delta(\epsilon-\bar\epsilon)+\delta(\epsilon+\bar\epsilon)]
\ee
with $\bar\epsilon$ a constant. The standard case corresponds to $q=1$ and displays a transition with diverging $\chi$ from an efficient ($H=0$) to a non efficient ($H>0$) regime when $n_s$ decreases below the critical value $n_s^*\simeq 4.15$.

\begin{figure}
\includegraphics[width=12cm]{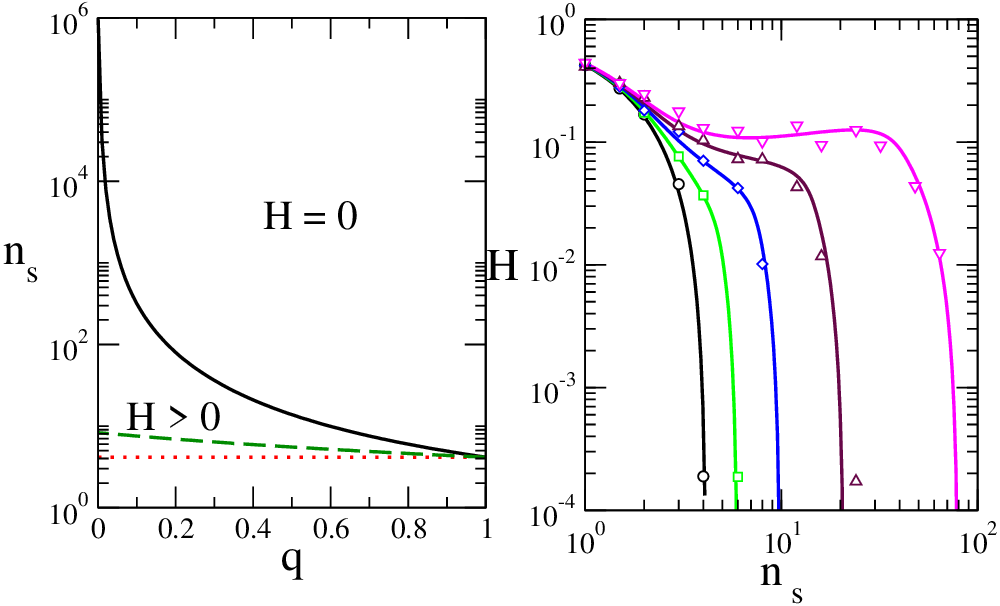}
\caption{\label{uno}(Colour on-line) {\bf Left panel:} phase diagram of the grand-canonical MG with heterogeneous incentives. The continuous line corresponds to the case where $R(\epsilon)$ is as in (\ref{ere}), the dotted horizontal line marks the transition point $n_s^*$ of the original model. {\bf Right panel:} $H$ vs $n_s$ for $q=1,0.8,0.6,0.4,0.2$ (left to right). Markers correspond to simulations with fixed $PN_s=32000$, averages over $100$ disorder samples.}
\end{figure}

In Fig. \ref{uno} we compare analytical predictions for different $q$ and $\bar\epsilon=0.25$. As expected, the phase diagram shows that the efficient phase shrinks as $q$ increases and, accordingly, the predictability vanishes.

Notice that the situation improves if an asymmetric distribution
$R(\epsilon)$ is considered. Indeed the dashed green line in
Fig. \ref{uno} represents the boundary between the efficient  and the
inefficient phase when
$R(\epsilon)=q\delta(\epsilon)+(1-q)\delta(\epsilon-1)$, showing that even
for very small but non-zero $q$ agents may wash out predictability from
the time series of bid imbalances. (Still, however, when $q=0$ no
efficient phase occurs.)

\section{Conclusions}

To summarize, we have studied the phase structure of batch MGs with mixed populations of optimally and sub-optimally learning agents, probing the robustness of the efficient phase against modifications of the agents' learning rules in the direction of increasing heterogeneity. Regimes with zero predictability  turn out to survive for any finite fraction of optimal agents. It would be interesting to know whether altering the information structure, that is taken to be fixed throughout the models discussed here, can modify this picture. For instance, a single agent (or a finite group of agents) with access to a more informative signal may alter this scenario or manage to take advantage from it. This direction is to our knowledge unexplored so far, with the possible exceptions of \cite{BMRZ,DG} and some studies along these lines would be in our opinion worthwhile.

\ack
This work is supported by an RCUK Fellowship (RCUK reference EP/E500048/1).



\section*{References}

\begin{thebibliography}{99}

\bibitem{book1}Challet D, Marsili M and Zhang YC 2005 Minority Games. (Oxford UP, Oxford, UK)
\bibitem{book2}Coolen ACC 2005 The mathematical theory of Minority Games. (Oxford UP, Oxford, UK)
\bibitem{rev}De Martino A and Marsili M 2006 J. Phys. A {\bf 39} R465
\bibitem{prl}Challet D, Marsili M and Zecchina R 2000 Phys. Rev. Lett. {\bf 84} 1824
\bibitem{hc}Heimel JAF and Coolen ACC 2001 Phys. Rev. E {\bf 63} 05612
\bibitem{impact1}De Martino A and Marsili M 2001 J. Phys. A {\bf 34} 2525
\bibitem{impact2}Heimel JAF and De Martino A 2001 J. Phys. A {\bf 34} L539
\bibitem{continuum} Marsili M, Challet D 2001 Phys. Rev. E {\bf 64} 056138  
\bibitem{GallSher05} Galla T and Sherrington D 2005  Eur. Phys. J. B {\bf 46} 153 
\bibitem{multistrat} De Martino A, Perez Castillo I and Sherrington D 2007 JSTAT P01006
\bibitem{gcmg}Challet D and Marsili M 2003 Phys. Rev. E {\bf 68} 036132
\bibitem{elfarol1} Challet D, Marsili M, Ottino G 2004 Physica A {\bf 332} 469
\bibitem{hetcl}De Sanctis L and Galla T 2006 JSTAT P12004
\bibitem{MCZ}Marsili M, Challet D and Zecchina R 2000 Physica A {\bf 522} 2000
\bibitem{dilut}Galla T 2005 JSTAT P01002 (2005)
\bibitem{finitemem}Challet D, De Martino A, Marsili M and Perez Castillo I 2006 JSTAT P03004
\bibitem{BMRZ}Berg J, Marsili M, Rustichini R and Zecchina R 2001 Quant. Finance {\bf 1} 203 (2001)
\bibitem{DG}De Martino A and Galla T 2005 JSTAT P08008













\end{thebibliography}
\end{document}